\documentclass[sigconf]{acmart}

\AtBeginDocument{%
  }

\copyrightyear{2026}
\acmYear{2026}
\setcopyright{cc}
\setcctype{by}
\acmConference[L@S '26] {Proceedings of the Thirteenth ACM Conference on Learning @ Scale}{June 29--July 3, 2026}{Seoul, Republic of Korea.}
\acmBooktitle{Proceedings of the Thirteenth ACM Conference on Learning @ Scale (L@S '26), June 29--July 3, 2026, Seoul, Republic of Korea}
\acmISBN{979-8-4007-2293-6/2026/06}
\acmDOI{10.1145/3774398.3811619}
\usepackage{graphicx}
\usepackage{subcaption}
\usepackage{booktabs}
\usepackage{siunitx}
\usepackage{balance}
\usepackage{hyperref}

\begin{document}

\title[Revisiting the Regularity of Student Learning Rate]{Revisiting the Regularity of Student Learning Rate:\\Sensitivity to Which Observations Are Included}

\author{Hansol Lee}
\affiliation{%
  \institution{Stanford University}
  \city{Stanford}
  \state{CA}
  \country{USA}
}
\email{hansol@stanford.edu}

\author{Guilherme Lichand}
\affiliation{%
  \institution{Stanford University}
  \city{Stanford}
  \state{CA}
  \country{USA}
}
\email{glichand@stanford.edu}

\author{Cristina Barnard}
\affiliation{%
  \institution{Stanford University}
  \city{Stanford}
  \state{CA}
  \country{USA}
}
\email{crisbg@stanford.edu}

\author{Lucas Klotz}
\affiliation{%
  \institution{University of California San Diego}
  \city{La Jolla}
  \state{CA}
  \country{USA}
}
\email{lucasiklotz@gmail.com}

\author{Candace Thille}
\affiliation{%
  \institution{Stanford University}
  \city{Stanford}
  \state{CA}
  \country{USA}
}
\email{cthille@stanford.edu}

\author{Yunsung Kim}
\affiliation{%
  \institution{Stanford University}
  \city{Stanford}
  \state{CA}
  \country{USA}
}
\email{yunsung@stanford.edu}

\author{Benjamin W. Domingue}
\affiliation{%
  \institution{Stanford University}
  \city{Stanford}
  \state{CA}
  \country{USA}
}
\email{bdomingu@stanford.edu}

\renewcommand{\shortauthors}{Hansol Lee et al.}

\begin{abstract}
Mixed-effects models fit to observational practice data are widely used in learning analytics to estimate student-level variation in initial knowledge and learning rate, and the resulting estimates increasingly inform substantive claims about learners. We examine whether such estimates can be read as properties of learners or whether they depend on choices about which observations the model is fit to. As a case study, we revisit the ``astonishing regularity'' reported by \citet{koedinger2023astonishing}: that students vary substantially in initial knowledge but much less in learning rate. The finding is based on fits of the individual Additive Factors Model (iAFM) to 27 educational datasets, and rests on a model-derived estimate of student-level learning-rate variation being small in absolute terms. We refit the same model on the same datasets under two specifications, each varying how much of each student's practice on a given skill is used in fitting. The estimate of student-level variation in initial knowledge stays approximately stable across both specifications. The estimate of student-level variation in learning rate does not: it inflates by a median of 118\% under one specification and is several times larger under the other. The same model, fit to the same data, returns substantially different estimates of how much students vary in learning rate depending on which observations are included. When estimates from mixed-effects models on observational practice data are used to support substantive claims about learners, sensitivity to such choices deserves a central place in how those estimates are reported and read.
\end{abstract}

\ccsdesc[500]{Applied computing~Education}
\ccsdesc[500]{Social and professional topics~Student assessment}
\ccsdesc[300]{Computing methodologies~Modeling and simulation}

\keywords{learning analytics; mixed-effects models; iAFM; learning rate; sensitivity analysis; practice length}

\maketitle

\section{Introduction}

Computer-based learning systems serving large student populations generate detailed records of practice: which tasks students attempted, in what order, and with what outcomes. Statistical models fit to these records are widely used to estimate properties of individual students, such as their initial knowledge of a skill and their rate of improvement with practice~\cite{cen2006learning, liu2017towards, pavlik2009performance, chi2011instructional}. The resulting estimates increasingly inform substantive claims about learners, including comparisons across subgroups~\cite{carvalho2024further}, evaluations of curricula~\cite{cen2007over}, and fairness audits of adaptive instruction~\cite{kizilcec2022algorithmic, baker2022algorithmic}. The conditions under which these estimates can be read straightforwardly as properties of learners, rather than as artifacts of how the underlying practice data is structured, therefore matter for what the field concludes.

A 2023 \textit{PNAS} study by Koedinger et al.~\cite{koedinger2023astonishing} fit a mixed-effects model called the individual Additive Factors Model (iAFM)~\cite{liu2017towards} to 27 educational datasets, comprising 1.3 million practice interactions from nearly 7,000 learners, and reported an ``astonishing regularity'' in student learning rates: students vary substantially in initial knowledge but learn at remarkably similar rates with practice. Across mathematics, language, science, and computer science, in elementary through college contexts, the authors interpret the regularity as evidence that achievement gaps reflect differences in prior preparation rather than differences in learning efficiency. The finding has been read as identifying a stable property of learners across a wide range of educational settings, and recent replication on data from over 15,000 students in the MATHia intelligent tutoring system~\cite{simpson2024replicating} has strengthened this reading.

The regularity claim rests on two estimates the model produces from each dataset. Variability in student learning rates is small in absolute terms: across the 27 datasets, the typical gap between students at the 25th and 75th percentiles is roughly 1 percentage point of accuracy per practice attempt. Variability in student initial knowledge is, by comparison, substantial: the same percentile gap in initial accuracy is roughly 20 percentage points. Together these support the interpretation that students differ substantially more in where they start than in how fast they learn. The model produces these estimates by fitting each student a single linear slope as a function of practice opportunities on each skill, and pooling slopes across students within each dataset.

The data the model is fit to is observational: students contribute very different numbers of practice attempts on each skill, with no experimental control over how much each student contributes. We find that across the datasets analyzed in \citet{koedinger2023astonishing}, the number of practice attempts a student contributes on a given skill ranges from a handful to several hundred within a dataset. Each per-student learning rate is therefore a linear summary of the student's trajectory fit through whatever range of opportunities the student contributed on each skill, and the pooled estimate of how much students differ in learning rate is a summary across slopes fit through different ranges. The original analysis in \citet{koedinger2023astonishing} includes supporting checks that establish the model can detect such heterogeneity when it is present in a fixed dataset; however, these checks do not address whether the heterogeneity it estimates depends on which practice attempts each per-student slope is fit through.

We take up this empirical question on the same 26 datasets. We replicate the iAFM under the original specification, then refit it under two specifications that change which observations enter the fit. The first caps each student's practice on each skill at the first ten attempts, equalizing the range of opportunities each per-student slope is fit through. The second, on the 9 of 26 datasets where data permit, fits the model separately on practice sequences of ten or fewer attempts and on those of more than ten attempts, and compares the resulting estimates. Across both, we find that the spread of student initial knowledge stays approximately stable, while the spread of student learning rates and the population-average learning rate shift substantially; in several of the short-sequence fits the former is many times its full-data value. The estimate the regularity claim rests on depends substantially on which observations enter the fit.

The specific finding above is that the regularity reported by \citet{koedinger2023astonishing} is sensitive to choices about which practice attempts the model is fit on, in ways the original analysis does not examine. More generally, the case illustrates that estimates of per-student parameters from student models fit to observational practice data depend on how that practice is distributed across students. As such estimates are increasingly used to draw substantive conclusions about learners, curricula, and adaptive instruction at scale, the structure of the data they are estimated from is worth examining alongside the models themselves.

\section{Related Work}

This section situates the paper in three areas of prior work: the model family the iAFM belongs to and the line of work the regularity finding sits within, what has been documented about how practice accumulates in observational learning data, and what is known about the reliability and sensitivity of estimates that student models produce from such data.

\subsection{The Additive Factors Model and the Regularity Finding}

The iAFM belongs to a family of mixed-effects logistic regression models for student performance on practice tasks. The Additive Factors Model (AFM)~\citep{cen2006learning} estimates an overall learning rate and per-skill effects from sequences of student attempts and has been used widely in learning analytics for evaluating cognitive models of skill~\citep{cen2006learning}, identifying skills given too much or too little practice~\citep{cen2007over}, and comparing skill models against learning data. Variants in this family address specific modeling concerns: Performance Factors Analysis~\citep{pavlik2009performance} distinguishes successes from failures in the practice history, and Instructional Factors Analysis~\citep{chi2011instructional} accommodates instructional events other than practice attempts. The iAFM~\citep{liu2017towards} extends the AFM with per-student intercept and slope terms, enabling estimation of individual differences in initial knowledge and learning rate.

\citet{liu2017towards} introduced the iAFM on two datasets selected for long per-student sequences and observed that estimating individual-level parameters reliably requires sufficient per-student data. \citet{koedinger2023astonishing} subsequently applied the iAFM at scale to 27 datasets and reported the regularity finding: substantial across-student variation in initial knowledge, small variation in learning rate. The finding has been replicated on a corpus of over 15{,}000 students in the MATHia tutoring system~\citep{simpson2024replicating} and extended to subgroups defined by demographic, academic-proficiency, and motivational variables~\citep{carvalho2024further}, and is increasingly cited as a stable property of learners in computer-based educational settings.

\subsection{Practice Length in Observational Learning Data}
\label{sec:related-practice-length}

The data fit by AFM-family models comes from logs of intelligent tutoring systems, online courses, and educational games. In all these settings, the number of attempts each student contributes on each skill is shaped by interactions among system design, content sequencing, and student behavior, rather than assigned by a study protocol. Mastery-based tutors~\citep{block1976mastery, frick1990comparison} advance students who reach a competence threshold, so faster learners contribute fewer attempts. Adaptive sequencing in intelligent tutoring systems~\citep{vanlehn2006behavior, koedinger2013new} routes students toward additional practice on skills they perform poorly on. \citet{cen2007over} found, in the Cognitive Tutor geometry curriculum, that some skills were over-practiced relative to mastery and others under-practiced. In open-ended environments, practice control shifts toward the student, so depth reflects engagement choices~\citep{aleven2003help}. Student behaviors further shape depth: \citet{beck2013wheel} characterized wheel-spinning, where students continue practicing without improving, and \citet{baker2007modeling, baker2006adapting} characterized off-task behavior and gaming the system.

The recognition that how much each student practices in observational learning data is not independent of how the student is performing has prompted recurring concern about what AFM-family models conclude from such data. \citet{beck2013wheel} suggested that long sequences from wheel-spinning students likely depress estimates of average learning rate. \citet{murray2013revealing} found in the Cognitive Tutor that disaggregating apparently flat aggregate curves by student subpopulation revealed learning in about 70\% of skills, attributing the aggregation artifact in part to mastery-based exit. \citet{goutte2018learning} identified an attrition bias in AFM-based learning curves and proposed a correction; \citet{pelanek2018details} and \citet{effenberger2020exploration} described the same concern under the name ``mastery attrition bias'' and flagged it as one the AFM does not address. These contributions establish that the structure of practice in observational learning data can affect what AFM-family models conclude. They have largely been raised at the level of individual datasets or curricula; the question has not been examined at the scale of the regularity finding.

\subsection{Reliability and Sensitivity of Student-Model Estimates}

Several lines of work bear directly on whether estimates from student models can be read straightforwardly as properties of learners. \citet{kaeser2014different} showed that different parameter combinations in learning-curve models can produce nearly identical predictions, an identifiability concern that complicates inference about individual students. \citet{beck2007identifiability} raised the same concern earlier for Bayesian Knowledge Tracing. \citet{liu2015variations} examined heterogeneity in learning rate by classifying students on systematic residual error patterns, a line of work the iAFM was later developed to make more direct. \citet{galyardt2015move} argued that not all observations contribute equally to estimating learner knowledge: more recent observations are more informative than older ones, suggesting that which observations enter a student model is a substantive modeling choice rather than an incidental one. Together, this work establishes that estimates from student models depend on choices about model specification and data selection that are not always made explicit when the estimates are used.

A separate but related thread concerns the shape of the underlying learning trajectory. The AFM and iAFM use a linear function of practice opportunities. The cognitive psychology of skill acquisition has long held that learning curves are concave: \citet{newell1981mechanisms} established the power law of practice as the dominant account, \citet{heathcote2000powerlaw} argued that the power-law shape in aggregated curves is an artifact of averaging over learners whose individual curves are better described by exponentials, and \citet{evans2018refining} proposed a more flexible form. A linear specification is therefore a useful approximation rather than a calibrated model of the underlying trajectory, and what such an approximation returns can depend on the range of opportunities each line is fit through.

\section{Background}

In this section we provide an overview of \citet{koedinger2023astonishing} and the iAFM they fit, as needed to follow the analyses in Section~\ref{sec:methods}.

\subsection{Data}

The 27 datasets that \citet{koedinger2023astonishing} analyze come from the Pittsburgh Science of Learning Center DataShop~\citep{koedinger2010datashop}.\footnote{\url{http://pslcdatashop.web.cmu.edu}} They cover elementary through college contexts and span mathematics, language, science, and computer science. The technology behind each dataset is an intelligent tutoring system, online course, or educational game. Sample sizes range from approximately 40 to nearly 4{,}000 students.

Each dataset records, for each student, a chronological sequence of attempts at tasks. For each attempt the dataset stores the student and task identifiers, the outcome (correct on first try or not), and the knowledge components (KCs) the task is annotated as requiring. The KC annotation comes from a skill model the analysis selects for each dataset; \citet{koedinger2023astonishing} restrict attention to skill models that assign each task to a single KC.

The analytic unit of the iAFM is the \emph{(student, KC) pair}: the sequence of attempts a particular student has made at tasks involving a particular skill, taken in chronological order. The number of attempts in a pair is not assigned by a study protocol; it emerges from how the educational technology routes students through content and from how each student engages with that content, and it varies substantially across pairs within every dataset.

\subsection{The iAFM}
\label{sec:bg-iafm}

The iAFM~\citep{liu2017towards} models each student's performance on a given skill as a straight line on the log-odds scale: an intercept that represents how well the student starts out on that skill, and a slope that represents how much each additional practice opportunity improves performance. Both the intercept and the slope are allowed to vary by student (some students start higher than others, some learn faster than others) and by skill (some skills are easier than others, some are learned faster than others). The model writes performance on a given task as the sum of these student-level and skill-level contributions.

Concretely, the iAFM predicts the log-odds that student $i$ answers task $j$ correctly on first try as
\begin{equation}
\label{eq:iafm}
\text{logit}(p_{ij}) = \underbrace{(\theta + \theta_i + \beta_{k(j)})}_{\text{intercept}} + \underbrace{(\delta + \delta_i + \gamma_{k(j)})}_{\text{slope}} \cdot T_{ik(j)},
\end{equation}
where $k(j)$ is the KC for task $j$ and $T_{ik(j)}$ is the number of prior opportunities student $i$ has had on KC $k(j)$ before this attempt. The intercept decomposes into a global baseline ($\theta$), a student-level deviation ($\theta_i$), and a KC-level deviation ($\beta_{k(j)}$). The slope decomposes the same way: a global learning rate $\delta$ plus per-student and per-KC deviations $\delta_i$ and $\gamma_{k(j)}$.

The global parameters $\theta$ and $\delta$ are \emph{fixed effects}: the model estimates one value of each for the whole dataset, capturing the typical student's initial-knowledge baseline and typical learning rate. The student- and skill-level deviations are \emph{random effects}: rather than estimating each student's $\theta_i$ and $\delta_i$ independently, the iAFM treats them as draws from a population distribution shared across students and estimates the parameters of that distribution. Specifically, $\theta_i$ and $\delta_i$ are modeled as having mean zero and variances $\sigma^2_\theta$ and $\sigma^2_\delta$ across the student population. These two variances are estimated by the fit, and their square roots $\widehat{\sigma}_\theta$ and $\widehat{\sigma}_\delta$ describe how much students vary from one another in initial knowledge and in learning rate. The skill-level deviations $\beta_k$ and $\gamma_k$ are modeled analogously, with their own variances.

The random-effects setup is what makes per-student estimation tractable when each student contributes limited data: the model borrows strength from the population to inform each student's value. Point estimates of the per-student deviations $\theta_i$ and $\delta_i$, written $\widehat{\theta}_i$ and $\widehat{\delta}_i$, are obtained as \emph{best linear unbiased predictors} (BLUPs). The borrowing of strength has a side effect: BLUPs are shrunk toward zero by an amount that decreases with how much data the student contributes, so a student with many observations has a BLUP close to what fitting that student alone would produce, while a student with few observations has a BLUP pulled heavily toward the population mean.

\subsection{The Regularity Finding}

Across the 27 datasets, \citet{koedinger2023astonishing} report three findings about the fitted iAFM.

\subsubsection{Typical values}

The median across datasets of $\theta$ corresponds to about 65\% accuracy on a typical KC at the first opportunity, indicating that the typical student is not at mastery at the start of practice. The median across datasets of $\delta$ is 0.09 log odds per opportunity, equivalent to a typical student requiring about 7.24 opportunities of practice to reach 80\% accuracy on a typical KC.

\subsubsection{Variation in initial knowledge is large}

Students vary substantially around the typical initial knowledge. \citet{koedinger2023astonishing} summarize student-level variation using both the variance of the BLUPs and their interquartile range (IQR), preferring IQR for its reduced dependence on distributional assumptions. The median across datasets of IQR$\{\widehat{\theta}_i\}$ is 0.830 log odds. Translated into mastery: a student at the 25th percentile of $\widehat{\theta}_i$ requires about 13 opportunities to reach 80\% accuracy on a typical KC, while a student at the 75th percentile requires about 4 (a 9-opportunity gap).

\subsubsection{Variation in learning rate is small in absolute terms}

In contrast, students appear similar in learning rate. The median across datasets of IQR$\{\widehat{\delta}_i\}$ is 0.018 log odds per opportunity, an order of magnitude smaller than the corresponding spread for KC-level learning rates (median IQR$\{\widehat{\gamma}_k\}$ of 0.132). Translated into mastery, a student at the 25th percentile of $\widehat{\delta}_i$ requires about 8 opportunities to reach 80\% accuracy, while a student at the 75th percentile requires about 7 (a 1-opportunity gap). The asymmetry between these two quantities is the regularity claim: students differ substantially in where they start, but similarly in how fast they learn.

\subsubsection{Supporting checks}

The original analysis includes several checks aimed at ruling out alternative explanations for the small spread in $\widehat{\delta}_i$. The iAFM outperforms the AFM (which omits the per-student slope $\delta_i$) by AIC in 21 of 27 datasets, indicating that some student-level slope variation is detectable in most datasets. Simulation studies confirm that the iAFM correctly recovers per-student slope variation in synthetic data when that variation is amplified. The variation in learning rate by KC is much larger than that by student, indicating that the model can detect slope heterogeneity in principle. Robustness checks across KC-model choice, across easier and harder KCs, and across grade-level subgroups all produce a similar pattern.

These checks establish that the iAFM, fit to a given dataset, can detect learning-rate variation when it is present in the data. They do not address whether the variation it estimates depends on the choice of which observations from each (student, KC) pair enter the fit, which is the question Section~\ref{sec:methods} takes up.

\section{Methods}
\label{sec:methods}

This section describes the empirical strategy by which we probe whether the iAFM's estimates of student-level variation depend on the structure of the practice data the model is fit to. We refit the iAFM under two specifications that vary that structure while holding the model and the fitting procedure fixed. The first specification holds the set of (student, KC) pairs constant and caps each pair at its first ten attempts. The second specification holds each pair's data intact and varies which pairs enter the fit, comparing fits on short and long pairs separately.

\subsection{Data and Model Fitting}
\label{sec:methods-data}

We use the same KC models (the assignments of skills to problems) and the same datasets as those reported in Table~S3 of \citet{koedinger2023astonishing}. Of the 27 datasets, 26 were accessible to us; dataset ds4555 was restricted and could not be obtained. Observations with missing KC labels or responses coded as unknown were removed. This preprocessing yielded observation counts identical to those in the original study for 25 of the 26 datasets; the only discrepancy occurred in ds392, where 154 additional observations were excluded under the same criteria. Following the original analysis, we retained only KCs with data from at least 10 students and a minimum of two practice opportunities. After preprocessing, the 26 datasets contain a combined 1.3 million attempts from approximately 7{,}000 students. We refer to the number of attempts in a (student, KC) pair as its \emph{pair length}, written $L$.

We fit the iAFM using the \texttt{lme4} package in R~\citep{bates2015fitting} with the \texttt{optimx} optimizer and the \texttt{nlminb} method, matching the configuration described in the original supplementary material. Four datasets (ds271, ds406, ds445, ds1899) yield singular full-data fits, in which a variance-component estimate hits zero or a correlation between random effects hits $\pm 1$. Singular fits indicate that the data are too sparse for one of the random-effects parameters to be reliably distinguished from a simpler model. We flag these where they bear on subsequent analyses.

\subsection{Outcome Measures}
\label{sec:methods-outcomes}

From each fit we extract three quantities: $\widehat{\sigma}_\theta$, $\widehat{\sigma}_\delta$, and $\widehat{\delta}$. The first two are the variance-component estimates introduced in Section~\ref{sec:bg-iafm}: the model's estimates of how much students vary in initial knowledge and learning rate, respectively. The third is the population-average learning rate, the fixed-effect estimate of $\delta$. For each specification, we compare these three quantities between the fit on the full data and the fit on the modified data, and we report within-dataset percent changes along with the count of datasets that move in a consistent direction.

\begin{figure}[t]
  \centering
  \includegraphics[width=\linewidth]{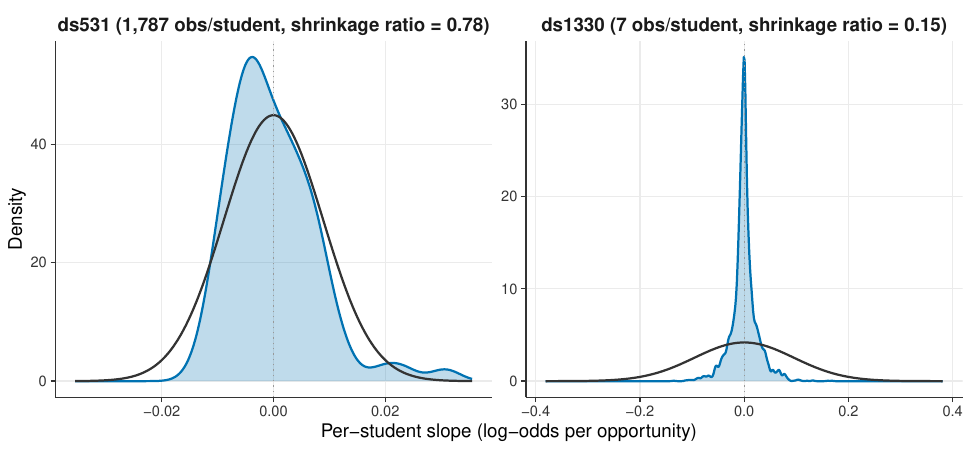}
  \caption{Empirical distribution of student-slope BLUPs (blue) and the Gaussian distribution $\mathcal{N}(0, \widehat{\sigma}_\delta^2)$ implied by the variance-component estimate from the same fit (black), for two datasets representing the extremes of per-student data density. The two distributions align when per-student data is dense (ds531, left, median 1{,}787 observations per student) and diverge sharply when it is sparse (ds1330, right, median 7 observations per student): in the sparse case, the BLUPs are compressed toward zero by shrinkage in a way that $\widehat{\sigma}_\delta$ is not. The shrinkage ratio noted in each panel header is the ratio of the empirical IQR of the BLUPs to the IQR implied by $\widehat{\sigma}_\delta$ under the model's Gaussian random-effects assumption ($\mathrm{IQR}\{\widehat{\delta}_i\} / (1.349 \cdot \widehat{\sigma}_\delta)$); values near 1 indicate little shrinkage, values near 0 indicate heavy shrinkage.}
  \label{fig:blup-vs-gaussian}
\end{figure}

\citet{koedinger2023astonishing} report a different summary of student-level spread: IQR$\{\widehat{\delta}_i\}$, the interquartile range of the BLUPs. Throughout the analyses, we report variance-component estimates rather than BLUP-IQRs. On the full-data fits the two summaries are close in absolute terms (median $\widehat{\sigma}_\delta$ of 0.025 versus median IQR$\{\widehat{\delta}_i\}$ of 0.020 across the 26 datasets), but they answer different questions and respond differently to the refits that follow. BLUPs are shrunk toward the population mean by an amount that depends on how much data each student contributes; our refits change per-student data density. A BLUP-based summary would therefore shift under our refits in part because density changes, not only because the population spread the model estimates changes. The variance-component estimate $\widehat{\sigma}_\delta$ is a property of the fitted distribution and is not subject to this density-dependent shrinkage. As an illustration, Figure~\ref{fig:blup-vs-gaussian} shows the divergence on two datasets: in ds531, where per-student data is dense, the BLUP distribution closely matches the Gaussian implied by $\widehat{\sigma}_\delta$; in ds1330, where it is sparse, the BLUPs are heavily compressed toward zero relative to that Gaussian.

\subsection{Truncation at Depth Ten}
\label{sec:methods-truncation}

The first specification caps each (student, KC) pair at its first ten attempts in chronological order. Pairs with ten or fewer attempts are unchanged. Pairs with more than ten attempts have their later attempts dropped, retaining only the first ten. The set of students and KCs that contribute to the fit is unchanged from the full-data fit. The iAFM is refit on the truncated data using the same procedure.

The motivation is that, in the full-data fit, the range of opportunities each per-student slope is fit through varies substantially across students within a dataset, from a handful to several hundred. Truncating at ten holds this range close to constant across students who reach the cap. Ten is also close to the median number of opportunities to mastery (7.24) reported in the original analysis (with most datasets in the range 5--15), so each per-student slope is fit through the practice region over which most students in the original analysis would have reached mastery.

Six datasets are excluded for one of two reasons. Three (ds115, ds271, ds1899) contain no pairs longer than ten attempts and are unaffected by truncation. Three more (ds406, ds445, ds562) yield singular fits under the full-data specification, the truncated specification, or both. The refit applies to the remaining 20 datasets.

\subsection{Pattern-Mixture Stratification}
\label{sec:methods-pattern-mixture}

The truncation specification holds the set of pairs constant and varies what each pair contributes within the fit. The second specification varies which pairs enter the fit. We fit the iAFM separately on two subsets of each dataset: pairs with ten or fewer attempts (the short-pair stratum) and pairs with more than ten attempts (the long-pair stratum). Within each stratum, all attempts in each retained pair are used. The within-stratum estimates are not pooled; we compare them directly.

The motivation is that the processes shaping how much students practice, reviewed in Section~\ref{sec:related-practice-length}, give reason to expect that short and long pairs sample different mixtures of underlying learning dynamics. Whether they do, in this corpus, is an empirical question the stratified comparison speaks to. Stratification by observed sequence length is the form pattern-mixture analysis~\cite{little1993pattern} takes in the longitudinal-data literature; we adopt the technique without committing to the missing-data interpretation that motivates it there.

This specification is feasible only when both strata contain enough data for the iAFM to produce non-singular stratum-level fits. Nine of the 26 datasets meet this criterion (ds99, ds392, ds406, ds447, ds564, ds565, ds566, ds567, ds1007); we report results for these nine. ds406 is included here even though it is excluded from the truncation specification above, since its full-data fit is singular but its stratum-level fits are not.

\section{Results}

This section reports the results of the analyses described in Section~\ref{sec:methods}. We first establish that our reimplementation reproduces the values reported by \citet{koedinger2023astonishing}. We then describe the distribution of practice length across the 26 datasets, since this distribution determines what each refit can and cannot do. Finally, we report the truncation results on the 20 datasets the specification applies to, and the pattern-mixture results on the 9 datasets where both strata support a non-singular fit.

\subsection{Replication}
\label{sec:results-replication}

\begin{figure}[t]
  \centering
  \includegraphics[width=\linewidth]{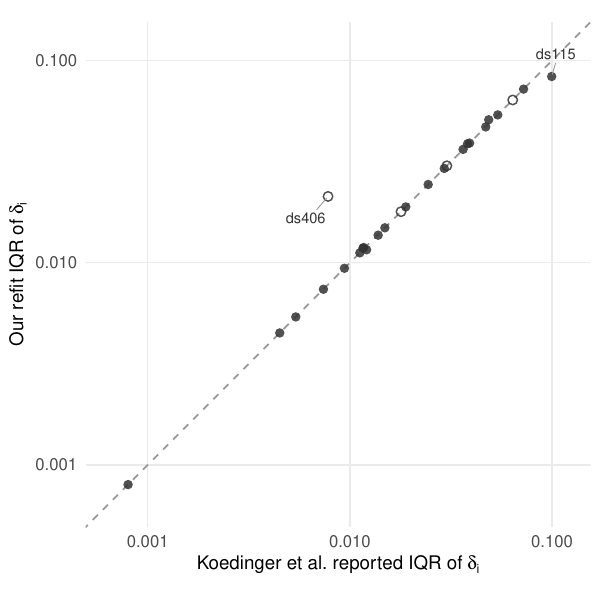}
  \caption{Replication of student learning-rate variability across 26 datasets. Each point shows our refit value compared to the value reported by \citet{koedinger2023astonishing}; the diagonal indicates perfect agreement. IQR$\{\widehat{\delta}_i\}$, the IQR of the BLUPs of the student-level slope random effects, agrees to within 0.003 for 24 of 26 datasets. The two exceptions are ds406 (where our fit is singular) and ds115 (where the dataset's maximum pair length is 4). Open circles indicate datasets with singular full-data fits.}
  \label{fig:replication}
\end{figure}

Our full-data refits reproduce the values reported by \citet{koedinger2023astonishing} closely. IQR$\{\widehat{\delta}_i\}$ agrees with the original to within 0.003 for 24 of 26 datasets; fixed-effect estimates agree to within 0.01 for 24 of 26 datasets (Figure~\ref{fig:replication}). The two discrepancies are ds406 (ours 0.021, theirs 0.008, with our fit singular) and ds115 (ours 0.084, theirs 0.100, on a dataset where the maximum pair length is only 4). Across the 26 datasets, the median IQR$\{\widehat{\delta}_i\}$ is 0.020 (closely matching the 0.018 originally reported), compared to a median IQR$\{\widehat{\gamma}_k\}$ of 0.136 for KC-level slopes and a median IQR$\{\widehat{\theta}_i\}$ of 0.797 for student-level intercepts. The reported pattern is reproduced: large variability across students in initial knowledge, small variability in learning rate.

\subsection{Distribution of Practice Length Across the Corpus}
\label{sec:distributions}

\begin{table*}[t]
\centering
\small
\caption{Per-dataset characteristics for the 26 datasets in our analysis, sorted by maximum pair length. Counts reflect our analytic sample (after KC and student filtering following \citet{koedinger2023astonishing}). \emph{Grade}: approximate grade level (Elem: upper elementary; Mid: middle school; HS: high school; Col: college). \emph{N$_{\text{stu}}$}: number of students. \emph{N$_{\text{KC}}$}: number of knowledge components. \emph{Obs}: total number of observations. \emph{Pair$_{\text{med}}$}, \emph{Pair$_{\text{max}}$}: median and maximum pair length, where a pair is a (student, KC) combination. \emph{Obs/stu}: median observations per student. \emph{\%pair$>$10}: percent of pairs with length exceeding 10 opportunities. \emph{\%obs$>$10}: percent of observations in pairs of length exceeding 10 opportunities. Datasets below the rule (ds1899, ds271, ds115) have no pairs exceeding length 10 and are not included in the truncation analysis (\S\ref{sec:truncation}).}
\label{tab:corpus}
\begin{tabular}{llllrrrrrrrr}
\toprule
Dataset & Domain & Edtech & Grade & N$_{\text{stu}}$ & N$_{\text{KC}}$ & Obs & Pair$_{\text{med}}$ & Pair$_{\text{max}}$ & Obs/stu & \%pair$>$10 & \%obs$>$10 \\
\midrule
ds531  & Language    & Tutor  & Col    & 77      & 68  & 117{,}297 & 15 & 335 & 1{,}787 & 89\% & 97\% \\
ds1935 & STEM        & Online & Col    & 39      & 17  & 8{,}601   & 5  & 326 & 253     & 16\% & 80\% \\
ds564  & Fractions   & Tutor  & Elem   & 73      & 60  & 66{,}342  & 8  & 201 & 906     & 44\% & 80\% \\
ds566  & Fractions   & Tutor  & Elem   & 58      & 58  & 63{,}907  & 8  & 189 & 1{,}098 & 32\% & 82\% \\
ds447  & Language    & Tutor  & Col    & 161     & 46  & 92{,}067  & 12 & 188 & 549     & 57\% & 79\% \\
ds99   & Geometry    & ITS    & HS     & 51      & 39  & 17{,}419  & 7  & 174 & 335     & 30\% & 73\% \\
ds253  & Geometry    & ITS    & HS     & 41      & 22  & 14{,}875  & 13 & 124 & 368     & 55\% & 89\% \\
ds392  & Geometry    & ITS    & Mid    & 123     & 38  & 41{,}602  & 8  & 113 & 337     & 30\% & 59\% \\
ds308  & Statistics  & Online & Col    & 52      & 9   & 4{,}152   & 3  & 99  & 90      & 14\% & 77\% \\
ds1007 & CompSci     & ITS    & Col    & 49      & 4   & 5{,}063   & 21 & 87  & 91      & 79\% & 95\% \\
ds1943 & Geometry    & ITS    & Mid    & 217     & 63  & 136{,}336 & 8  & 75  & 703     & 39\% & 72\% \\
ds406  & CompSci     & Online & Col    & 764     & 69  & 440{,}235 & 7  & 68  & 584     & 33\% & 70\% \\
ds563  & Fractions   & Tutor  & Elem   & 64      & 54  & 55{,}407  & 11 & 56  & 864     & 50\% & 79\% \\
ds565  & Fractions   & Tutor  & Elem   & 61      & 78  & 57{,}948  & 8  & 53  & 945     & 41\% & 71\% \\
ds1387 & Fractions   & ITS    & Elem   & 70      & 31  & 3{,}910   & 3  & 40  & 55      &  4\% & 19\% \\
ds104  & Physics     & ITS    & Col    & 104     & 12  & 6{,}019   & 3  & 38  & 62      & 13\% & 54\% \\
ds562  & Fractions   & Tutor  & Elem   & 63      & 97  & 48{,}404  & 6  & 36  & 769     & 22\% & 52\% \\
ds567  & Fractions   & Tutor  & Elem   & 59      & 194 & 45{,}955  & 3  & 30  & 774     &  4\% & 12\% \\
ds1980 & English     & Tutor  & Col    & 120     & 16  & 6{,}592   & 7  & 26  & 56      & 13\% & 35\% \\
ds445  & Fractions   & Game   & Elem   & 51      & 20  & 4{,}275   & 2  & 20  & 84      & 10\% & 38\% \\
ds394  & English     & Tutor  & Col    & 97      & 13  & 5{,}773   & 5  & 15  & 60      & 10\% & 24\% \\
ds1330 & Algebra     & Tutor  & Mid    & 3{,}819 & 23  & 38{,}938  & 2  & 14  & 7       & $<$1\% & $<$1\% \\
ds372  & English     & Tutor  & Adult  & 99      & 19  & 7{,}128   & 5  & 12  & 72      &  9\% & 17\% \\
\midrule
ds1899 & Algebra     & Tutor  & Mid    & 302     & 6   & 4{,}930   & 2  & 10  & 16      &  0\% &  0\% \\
ds271  & Algebra     & Tutor  & Mid    & 69      & 6   & 1{,}103   & 2  & 8   & 16      &  0\% &  0\% \\
ds115  & Chinese     & ITS    & Col    & 72      & 166 & 16{,}270  & 2  & 4   & 226     &  0\% &  0\% \\
\bottomrule
\end{tabular}
\end{table*}

Table~\ref{tab:corpus} summarizes the 26 datasets. The corpus spans elementary through college contexts; covers mathematics, language, science, and computer science; and includes intelligent tutoring systems, online courses, and educational games. Sample sizes range from 41 to 3{,}819 students, from 4 to 194 KCs, and from approximately 1{,}100 to 440{,}000 observations per dataset.

Two features of the corpus matter for the analyses that follow. First, practice length is highly imbalanced across (student, KC) pairs within each dataset. The median pair length ranges across datasets from 2 to 21; the maximum pair length ranges from 4 to 335. Within individual datasets, the gap between typical and maximum pair length is often large: ds1935 has a median pair length of 5 but a maximum of 326, and ds308 has a median of 3 but a maximum of 99. In most datasets, most pairs are short, with a small minority running much longer.

Second, the imbalance is consequential because long pairs contribute disproportionately to total observations. In ds1935, 15.9\% of pairs exceed length 10 but account for 80.0\% of observations in the dataset; in ds308, 13.9\% of pairs account for 77.0\% of observations. The iAFM weights observations uniformly in its likelihood, so the small set of long pairs receives weight in the fit far out of proportion to the share of pairs they represent. Across the 23 datasets with at least one pair longer than 10, the share of observations contributed by such pairs ranges from 0.8\% (ds1330) to 97.0\% (ds531).

\begin{figure}[t]
  \centering
  \includegraphics[width=\linewidth]{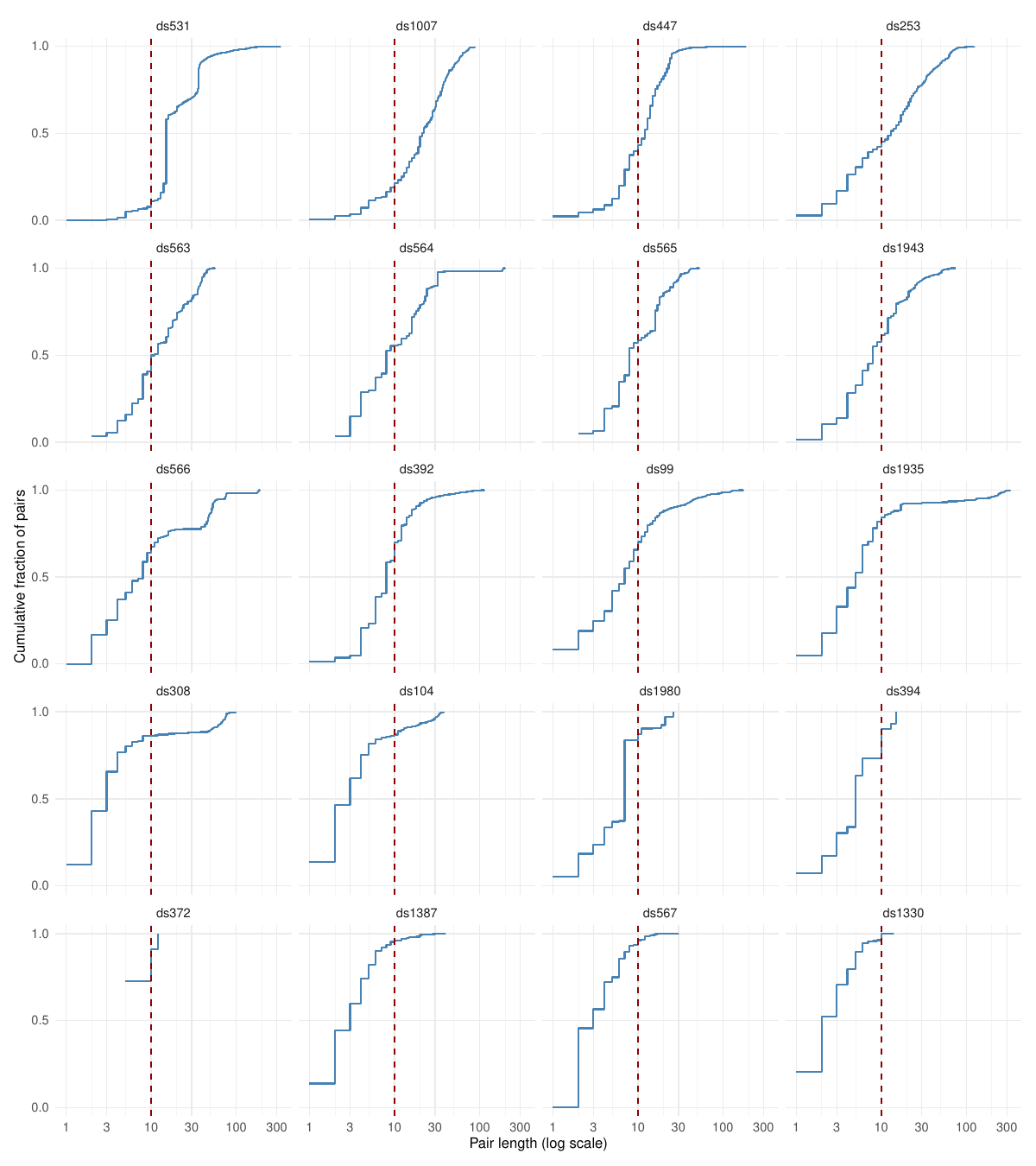}
  \caption{Empirical cumulative distribution of (student, KC) pair lengths for the 20 datasets included in the truncation analysis (\S\ref{sec:truncation}). Datasets are ordered by the fraction of pairs with $L \leq 10$, from lowest (top-left) to highest (bottom-right). The dashed red line marks $L = 10$, the truncation threshold.}
  \label{fig:pair-length-ecdf}
\end{figure}

Figure~\ref{fig:pair-length-ecdf} shows the empirical cumulative distribution of pair lengths for the 20 datasets the truncation specification applies to. Datasets are ordered by the fraction of pairs at or below 10.

\subsection{Truncation Refit}
\label{sec:truncation}

\begin{figure*}[t]
  \centering
  \includegraphics[width=\linewidth]{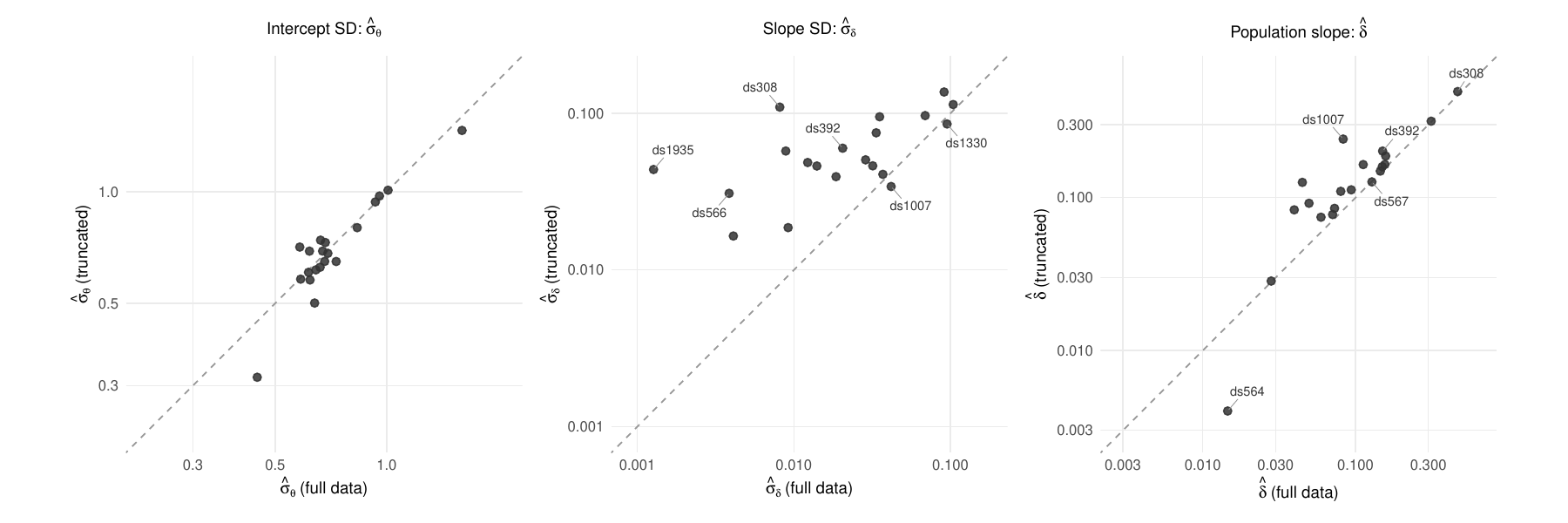}
  \caption{iAFM parameter estimates under full and truncated ($L = 10$) data, one point per dataset (20 datasets). Left: student-intercept SD $\widehat{\sigma}_\theta$, approximately stable across the comparison. Middle: student-slope SD $\widehat{\sigma}_\delta$, inflating in 18 of 20 datasets. Right: population-average slope $\widehat{\delta}$, steepening in 18 of 20 datasets. All axes on log scale. Dashed line indicates equality.}
  \label{fig:truncation}
\end{figure*}

Figure~\ref{fig:truncation} compares full-data and truncated fits on the 20 eligible datasets, with one panel per parameter. The left panel shows the student-intercept SD $\widehat{\sigma}_\theta$: points lie tightly along the diagonal, indicating that truncation barely affects this estimate. The median within-dataset percent change is $-1.5\%$ (IQR $-5.7\%$ to $+2.1\%$). The middle panel shows the student-slope SD $\widehat{\sigma}_\delta$: 18 of 20 points sit clearly above the diagonal, with several (ds308, ds1935, ds566) far up and to the left where small full-data baselines combine with substantially larger truncated values. The median within-dataset percent change is $+118\%$ (IQR $+44\%$ to $+297\%$). The right panel shows the population-average slope $\widehat{\delta}$: 18 of 20 points again sit above the diagonal, and the median rises from 0.088 to 0.126 (median within-dataset increase $+18\%$, IQR $+5\%$ to $+40\%$). The visual contrast between the left panel and the other two is the headline result: $\widehat{\sigma}_\theta$ is stable under truncation while $\widehat{\sigma}_\delta$ and $\widehat{\delta}$ are not.

The magnitude of $\widehat{\sigma}_\delta$ inflation varies across the 20 datasets and is associated with how much data truncation removes. The within-dataset percent change in $\widehat{\sigma}_\delta$ correlates positively with the fraction of observations truncation drops (Spearman $\rho = 0.68$). Datasets where truncation removes more observations show, on average, larger inflation; the residual variation indicates that other features of the practice-length distribution and the underlying curves also contribute. Relative changes are also sensitive to the magnitude of the full-data baseline: in datasets where the full-data $\widehat{\sigma}_\delta$ is near zero, the percent change can be very large for a modest absolute shift.

Two datasets sit on or below the diagonal in the middle panel and are the exceptions to the $\widehat{\sigma}_\delta$ inflation pattern. They sit at opposite ends of the practice-length distribution. In ds1330, with median pair length 2 and only 0.2\% of pairs affected by the cap, truncation operates on almost no observations. In ds1007, with median pair length 21 (the longest in the corpus), most pairs are long enough that even after truncation each per-student slope is fit through a similar range. The inflation pattern is concentrated among datasets in the middle of the length distribution, where truncation removes a substantial proportion of observations while still leaving enough data per pair for the model to fit. In the right panel, the most visible exception is ds564, well below the diagonal; ds567 also sits just below the diagonal but moves by less than 1\%.

\subsection{Pattern-Mixture Stratification}
\label{sec:pattern-mixture}

\begin{table}[t]
\centering
\small
\caption{iAFM parameter estimates under pattern-mixture stratification, for the 9 datasets with non-singular fits in both strata. Each dataset is fit separately on its short pairs ($L \leq 10$) and long pairs ($L > 10$). $\widehat{\sigma}_\theta$: student-intercept SD. $\widehat{\sigma}_\delta$: student-slope SD. $\widehat{\delta}$: population-average slope.}
\label{tab:pattern-mixture}
\begin{tabular}{lrrrrrr}
\toprule
& \multicolumn{2}{c}{$\widehat{\sigma}_\theta$} & \multicolumn{2}{c}{$\widehat{\sigma}_\delta$} & \multicolumn{2}{c}{$\widehat{\delta}$} \\
\cmidrule(lr){2-3} \cmidrule(lr){4-5} \cmidrule(lr){6-7}
Dataset & Short & Long & Short & Long & Short & Long \\
\midrule
ds99   & 0.638 & 0.758 & 0.103 & 0.014 & 0.204 & 0.080 \\
ds392  & 0.674 & 0.439 & 0.095 & 0.013 & 0.284 & 0.127 \\
ds406  & 0.327 & 0.301 & 0.040 & 0.012 & 0.157 & 0.013 \\
ds447  & 0.685 & 0.689 & 0.104 & 0.030 & 0.109 & 0.081 \\
ds564  & 0.768 & 0.665 & 0.037 & 0.004 & 0.011 & 0.003 \\
ds565  & 0.608 & 0.785 & 0.047 & 0.026 & 0.145 & 0.055 \\
ds566  & 0.654 & 0.610 & 0.016 & 0.005 & 0.142 & $-$0.001 \\
ds567  & 0.689 & 0.785 & 0.051 & 0.040 & 0.125 & 0.002 \\
ds1007 & 1.212 & 1.513 & 0.114 & 0.039 & 0.266 & 0.077 \\
\bottomrule
\end{tabular}
\end{table}

\begin{figure*}[t]
  \centering
  \includegraphics[width=\linewidth]{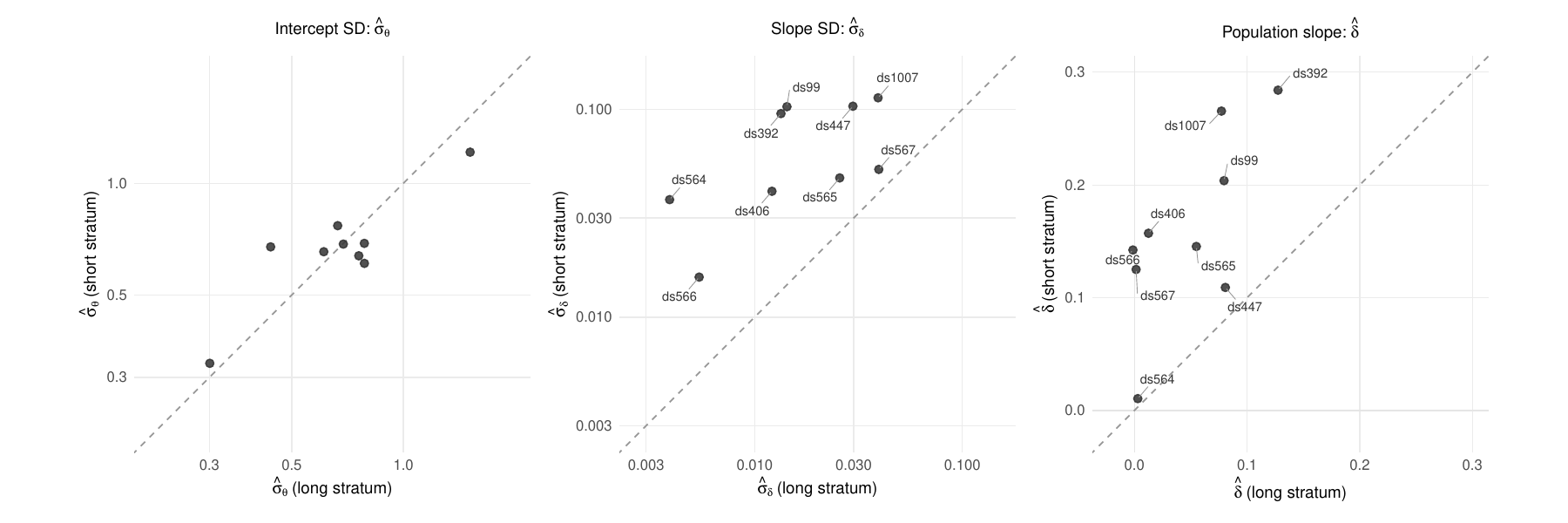}
  \caption{iAFM parameter estimates within the short-pair stratum ($L \leq 10$) and long-pair stratum ($L > 10$), one point per dataset (9 datasets with non-singular stratum-level fits). Left: student-intercept SD $\widehat{\sigma}_\theta$, approximately stable across strata. Middle: student-slope SD $\widehat{\sigma}_\delta$, substantially larger in the short stratum in 9 of 9 datasets. Right: population-average slope $\widehat{\delta}$, substantially steeper in the short stratum in 9 of 9 datasets. SD axes on log scale; $\widehat{\delta}$ on linear scale. Dashed line indicates equality.}
  \label{fig:pattern-mixture}
\end{figure*}

Table~\ref{tab:pattern-mixture} reports the iAFM's parameter estimates within each of the short-pair ($L \leq 10$) and long-pair ($L > 10$) strata for the 9 datasets where both strata support a non-singular fit; Figure~\ref{fig:pattern-mixture} visualizes the same comparison. The three panels show the same asymmetric pattern as the truncation analysis, in sharper form. In the left panel, the student-intercept SD $\widehat{\sigma}_\theta$ scatters around the diagonal: the median within-dataset percent difference (short minus long) is essentially $0\%$ (IQR $-16\%$ to $+9\%$), and the short-stratum value is higher in only 4 of 9 datasets. In the middle panel, the student-slope SD $\widehat{\sigma}_\delta$ sits above the diagonal for all 9 datasets; the median percent difference is $+233\%$ (IQR $+189\%$ to $+613\%$). In the right panel, the population-average slope $\widehat{\delta}$ sits above the diagonal for all 9 datasets, with several points (ds406, ds567) far from the diagonal where the long-stratum $\widehat{\delta}$ is near zero. The median absolute difference (short minus long) is 0.12 log-odds per opportunity, comparable to the typical full-data $\widehat{\delta}$ of 0.09 reported by \citet{koedinger2023astonishing}; we report the absolute difference rather than a percent change because of those near-zero long-stratum values.

The absolute magnitudes are substantial. In 8 of the 9 datasets, short-stratum $\widehat{\sigma}_\delta$ exceeds the original median IQR$\{\widehat{\delta}_i\}$ of 0.018 on which the regularity finding rests: 0.114 in ds1007, 0.104 in ds447, 0.103 in ds99, 0.095 in ds392, 0.051 in ds567, 0.047 in ds565, 0.040 in ds406, and 0.037 in ds564. The exception is ds566, whose short-stratum value (0.016) is comparable to the baseline but still more than three times its long-stratum value (0.005). ($\widehat{\sigma}_\delta$ and IQR$\{\widehat{\delta}_i\}$ are not the same summary, but both are on the same log-odds-per-opportunity scale and were close in the full-data fits; see Section~\ref{sec:methods-outcomes}.) The $\widehat{\delta}$ values diverge similarly across strata: in ds392, short-stratum $\widehat{\delta} = 0.284$ versus long-stratum $0.127$; in ds406, $0.157$ versus $0.013$; in ds566, $0.142$ versus essentially zero ($-0.001$). The same model, fit to the same dataset, produces population learning rates that differ by a factor of two or more in 7 of the 9 datasets, with the difference reaching an order of magnitude in ds406 and ds567.

\section{Discussion}
\label{sec:discussion}

Two estimates the iAFM produces from a dataset respond very differently to our refits. The estimate of how much students vary in initial knowledge remains approximately stable when we change which observations from each (student, KC) pair the model is fit to. The estimate of how much students vary in learning rate does not: it inflates by a median of 118\% under truncation at the first ten attempts, and is several times its full-data value in the short-pair stratum across all 9 datasets where the stratified comparison is feasible. The regularity finding rests on the second of these estimates being small in absolute terms; in our refits, that estimate is not stable.

The asymmetry has a structural source. The student-level intercept that the iAFM estimates is anchored at the first practice opportunity: every student contributes an observation at that opportunity on each pair they enter, and at $T_{ik(j)} = 0$ the model's predicted log-odds reduces to the intercept terms alone. The spread of intercepts is therefore identified directly from those observations, and refits that alter data from later opportunities leave it intact. The student-level slope is identified differently. Each per-student slope is a single line fit through the range of opportunities the student contributed on a given skill, and the pooled slope distribution aggregates slopes fit through different segments of the practice trajectory. When these ranges differ substantially across students within a dataset, the estimate of how much students vary in slope depends on which ranges enter the fit. Our refits change that mixture, and the estimate shifts accordingly.

Several mechanisms suggest that the composition of this mixture should matter. Cognitive psychology has long held that learning curves are concave~\citep{newell1981mechanisms, heathcote2000powerlaw, evans2018refining}; a linear approximation of a concave curve produces a steeper slope when fit through early opportunities and a flatter slope when fit through later ones, consistent with both the steepening of $\widehat{\delta}$ and the inflation of $\widehat{\sigma}_\delta$ we observe under truncation. How much each student practices in observational learning data is also not independent of how the student is performing: mastery-based exit, adaptive routing, wheel-spinning, and disengagement all couple practice length to performance~\citep{beck2013wheel, baker2006adapting, vanlehn2006behavior}. Prior work on AFM-family models has documented related concerns under the labels of attrition bias and mastery-attrition bias~\citep{murray2013revealing, goutte2018learning, pelanek2018details, effenberger2020exploration}; our results extend those concerns from aggregate AFM learning curves to the iAFM's student-level variance estimates. Identifiability limits at small per-student sample sizes~\citep{kaeser2014different, beck2007identifiability} may also contribute. The analyses here document the asymmetry without adjudicating among these mechanisms.

A related conceptual question bears on how learning rate should be measured. The iAFM's per-student slope is a well-defined quantity: an average slope on the log-odds scale over whatever practice the student contributed on each skill. Within a dataset, however, that window can vary substantially across students: one student's slope might be fit over 5 attempts and another's over 300, depending on how the system routed them and how they engaged. Across datasets, the typical window varies further with system design. A slope fit over the first 10 attempts is not the same quantity as one fit over the first 300, since additional practice past the point of mastery flattens the curve~\citep{anderson1982acquisition, fitts1967human}; comparing such slopes across students mixes information about how fast learners improve with information about how long they kept practicing. This is not a problem with the iAFM, which produces a well-defined quantity given the data it is fit to. The question is what that quantity corresponds to as a property of learners: what window of observed practice operationalizes ``learning rate'' when those windows differ this much across students. \citet{galyardt2015move} make a related point in a temporal frame, and the regularity finding puts the question into sharper relief.

These results do not establish that students learn at substantially different rates. The short-stratum and truncated fits are not privileged estimates of any underlying ``true'' student-level variation; they are different fits of the same approximating model to differently selected data. Whether learning rate varies more, less, or about as much across students as the regularity claim suggests is a substantive question this paper does not resolve. What our results do establish is that the published estimate of small slope variation warrants checking sensitivity to which observations enter the fit. The same logic applies to recent replications: \citet{simpson2024replicating} reproduced the regularity on a corpus of over 15{,}000 students in MATHia using the same iAFM specification on the same kind of observational data, and \citet{carvalho2024further} extended it to subgroups defined by demographic, academic-proficiency, and motivational variables. Both establish robustness across cohorts and subgroup splits; neither addresses the sensitivity we document, which would be expected to operate in those settings as well.

The regularity finding has been interpreted as evidence about a question with policy implications. \citet{koedinger2023astonishing} read the finding as suggesting that ``educational achievement gaps come from differences in learning opportunities and that better access to such opportunities can help close those gaps.'' Under this interpretation, learning rate is approximately uniform across students, and observed differences in attainment trace back to differences in cumulative exposure rather than to differences in how efficiently students learn from a given exposure.

Our findings do not refute this interpretation, but they qualify its quantitative basis. The interpretation depends on the spread of student learning rates being small as a property of learners. Under specifications that hold the practice range closer to constant across students, the spread of estimated student slopes is several times its full-data value and, in some datasets, comparable to or larger than the variation across knowledge components. Cumulative advantage offers an alternative interpretation compatible with both the original analysis and ours: students may start at different points and learn at different rates, with both differences compounding over cumulative exposure. Adjudicating between these interpretations requires evidence the existing analyses do not provide; the substantive question of how much students differ in learning rate, when learning conditions are held constant, remains open.

For learning-analytics studies that fit the iAFM and related student models to observational practice data, estimates of student-level variation should be reported alongside descriptions of the practice distribution from which they are derived: at minimum, the median pair length, the spread of pair lengths, and the share of observations contributed by long pairs. Sensitivity checks of the type reported here are simple enough to be adopted as routine diagnostics. The concern is more acute for studies that compare estimated learning rates across groups, conditions, or interventions. When practice exposure differs systematically across the groups being compared, for instance because the system routes struggling students to additional practice or because subgroups receive different scaffolds or completion rates, the practice-length distributions for those groups will differ in ways that may correlate with the parameters being compared. Reported differences in estimated learning rate or efficiency between conditions may then reflect differences in exposure rather than differences in learning. The concern extends to evaluations of tutoring systems and to fairness audits of adaptive instruction~\citep{kizilcec2022algorithmic, baker2022algorithmic}, where the populations under comparison often differ in precisely the way that produces the sensitivity we document. The broader principle, articulated by \citet{galyardt2015move}, is that decisions about which observations contribute to estimating learner properties warrant explicit attention rather than being left implicit in the modeling pipeline.

\subsection{Limitations and Future Directions}

The refits document sensitivity but do not test the mechanisms that produce it. Ten is one defensible truncation cap, close to the median number of opportunities to mastery the original analysis reports (7.24), but it is a single threshold. Three further analyses would help characterize what we observe. Sweeping across truncation thresholds, alongside dataset-level inspection of aggregate empirical learning curves, would document how the inflation varies with the cap and where curvature is concentrated within the practice range. Refitting the iAFM with nonlinear opportunity terms (logarithmic, saturating, or power-law) would isolate functional-form contributions from observation-process contributions to the asymmetric signature. Refitting on a random window of $L$ attempts from each pair, rather than the first $L$, would test whether the pattern is tied specifically to the early region of practice or to per-pair information reduction more generally.

The analyses are also bounded by what the available data support. The pattern-mixture refit is feasible only on the 9 of 26 datasets with adequate data in both strata, a subset selected on data adequacy rather than at random; whether the same pattern holds in the 17 datasets where the comparison is infeasible remains open. The analyses hold pair length fixed across datasets that differ substantially in the calendar time over which attempts were recorded, so the same nominal pair length represents qualitatively different learning situations across platforms; whether and how temporal structure matters for the estimates is a question we do not pursue here. The linear specification in practice opportunities at the core of the iAFM is shared across much of the AFM family~\citep{cen2006learning, pavlik2009performance, chi2011instructional}, and we would expect the asymmetric signature wherever a linear approximation is fit across heterogeneous practice ranges; whether the same pattern holds for nonlinear or non-AFM student models is a separate question. The quantities we track are also parameter-level summaries; how much of the observed movement translates to outcome-scale differences in predicted accuracy is not addressed here.

If how much each student practices proves informative about the underlying learning trajectory, a more general response is to model the observation process explicitly. Joint longitudinal and time-to-event models~\citep{rizopoulos2012joint, tsiatis2004joint, wu2012analysis}, selection models~\citep{diggle1994informative, wu1988estimation}, and pattern-mixture models~\citep{little1993pattern} fit the outcome trajectory and the observation process simultaneously. Adapting any of these to the iAFM requires modeling how the system and the student jointly determine when practice ends, which varies across the platforms in this corpus and would itself become an object of study. Until such work is undertaken, the simple refits we report here are immediately applicable; reporting them alongside the variance-component estimates in studies that fit iAFM-family models to observational practice data would substantially improve what readers can take from the resulting estimates.

\section{Conclusion}

The ``astonishing regularity'' in student learning rate is a finding about an estimate the iAFM produces from observational practice data. On the same 26 datasets, that estimate moves substantially under reasonable changes to which observations enter the fit, while the companion estimate of student initial-knowledge variation does not. The asymmetry points to a more general issue: estimates of student-level variation from mixed-effects models fit to observational practice data depend on choices about which observations to include that are not always made explicit. As such models continue to be fit at scale and used to inform educational practice, sensitivity to these choices deserves a central place in how the resulting estimates are reported and read.

\bibliographystyle{ACM-Reference-Format}
\balance
\bibliography{mybib}

\end{document}